\documentclass[twocolumn, twoside,prl]{revtex4}

\usepackage{amsmath}
\usepackage{amssymb}
\usepackage{latexsym}
\usepackage{revsymb}
\usepackage{epsfig}

\newcommand{\barr}{\begin{eqnarray}}
\newcommand{\earr}{\end{eqnarray}}
\newcommand{\ra}{\rangle}
\newcommand{\la}{\langle}
\newcommand{\beq}{\begin{equation}}
\newcommand{\eeq}{\end{equation}}

\begin{document}

\title{The classical limit of quantum optics: not what it seems at first sight}
\author{Yakir Aharonov$^{1,2}$}
\author{Alonso Botero$^{3}$}
\author{Shmuel Nussinov$^1$}
\author{Sandu Popescu$^{4}$}
\author{Jeff Tollaksen$^2$}
\author{Lev Vaidman$^{1,2}$}
\affiliation{$^1$ School of Physics and Astronomy, Tel Aviv University,  %
Tel Aviv, Israel} \affiliation{$^2$ Institute of Quantum Studies and Faculty of Physics, Chapman University,
1 University Drive, Orange, CA 92866, USA}
\affiliation{$^3$ Departamento de Fisica, Universidad de Los Andes, Bogota, Columbia}
\affiliation{$^4$ H.H.Wills Physics Laboratory, University of Bristol, %
 Tyndall Avenue, Bristol BS8 1TL, U.K.}

\begin{abstract}

What is light and how to describe it has always been a central subject in physics. As our understanding has
increased, so have our theories changed: Geometrical optics, wave optics and quantum optics are increasingly
sophisticated descriptions, each referring to a larger class of phenomena than its predecessor. But how exactly
are these theories related?  How and when wave optics reduces to geometric optics is a rather simple problem.
Similarly, how quantum optics reduces to wave optics has been considered to be a very simple business as well.
It's not so. As we show here the classical limit of quantum optics is a far more complicated issue; it is in
fact dramatically more involved and it requires a complete revision of all our intuitions. The revised
intuitions can then serve as a guide to finding novel quantum effects.

\end{abstract}

\maketitle

Inventing a new theory of nature requires, as R. Feynman said, ``imagination in a terrible strait-jacket''\cite{feynman}. Unlike the artist who need not obey constraints, as scientists we are not free to imagine whatever we
want - the new theory must obey a ``correspondence principle''. It must, obviously, give different predictions
from those of the old theory for some phenomena, but at the same time it must agree with the old theory in all the places
in which the old theory was already experimentally verified. For all those experiments the new theory must give
numerical results that are very similar to those of the old theory; the only acceptable difference must be
smaller than the precision of the measurements that seemed to confirm the old theory.

Insofar as the {\it numerical} predictions of a theory are concerned, the correspondence principle is fairly
obvious and straightforward. However, theories are not only mathematical devices for making numerical
predictions - they contain {\it concepts} that tell us a story of what the nature of physical reality is. And,
as Feynman also noted, even in situations when the numerical predictions of two theories are almost identical,
the concepts they involve may be completely different. Indeed, it is a fundamental conceptual difference
between, say, mass being an absolute constant or mass changing with the speed even when the speed is so low that
the change of mass is negligible.

Yet, and this is the point that concerns us here, although the concepts of the two theories are completely
different, and remain different even in the regime in which the two theories give almost identical numerical
predictions, it seems to be the case that the story told by the new theory relates in rather simple ways to the
story told by the old theory. For example, although mass changes with speed, at low speeds it changes only very
little - it has a certain simple continuity relation with the old story of absolute constant mass. It doesn't
have to be so - there is no need of any ``continuity'' at the conceptual level, only the numerical predictions
have to agree in the regime where the old theory was already verified - but the history of science seems to
suggest that this is always the case.

Even the classical limit of quantum mechanics seems to be rather benign - with increasing mass, wavepackets can
be taken narrower and narrower yet they will take longer and longer to spread, so particles can be better and
better localized. Wavepackets will then follow trajectories very close to the classical ones. As we show here
however, this benign behavior is only an illusion: we will present a situation (in the context of optics) where
the classical limit of quantum mechanics tells a conceptual story that is dramatically different from that of
classical physics.

An essential point to emphasize is that this whole discussion is not just restricted to interpretations. Quite the
opposite. Allowing us to have a better intuition is essential for finding new and interesting quantum effects
and may lead to new experiments and potential practical applications.

To start with, consider a beam of monochromatic light of frequency $\omega$ and high intensity $I$ impinging on
a mirror at an angle $\alpha$ as illustrated in Fig.1. Suppose the beam has unit area and the experiment lasts a
unit time. Classically, light is a wave and it carries a momentum proportional to its intensity. Upon reflecting
from the mirror, it gives the mirror in each unit of time a momentum kick \beq\delta p_M=2I\cos\alpha,\eeq equal to twice
the component of the momentum normal to the mirror (for simplicity, throughout of this paper we take the speed
of light $c=1$). Quantum mechanically the story is different. Light consist of photons, each carrying a
momentum $p=\hbar\omega$. The quantum state that best describes the physics of light in this ``classical''
regime is the so called ``coherent state'' \cite{sud,glau}, which is a superposition of different photon numbers with average
number $\overline n$ and spread $\Delta n=\sqrt{\overline n}$. All photons behave in an identical way: Each
photon bounces off the mirror and gives it a momentum kick, and the total average momentum imparted to the mirror
is $\overline n$ times the momentum given by each photon \beq\delta p_M =2{\overline n}\hbar \omega\cos
\alpha.\eeq Given that the intensity of light is $I={\overline n}\hbar \omega$, the momentum kick
calculated quantum mechanically is identical to the classical one.

\begin{figure}
\epsfig{file=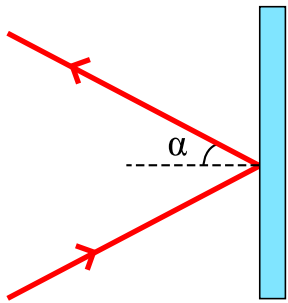, scale=0.25} \caption{Light reflected on a mirror}
\end{figure}

Although the classical and quantum stories are different, they seem nevertheless quite close to each other.
Following this paradigmatic example, it is tempting to think that the classical-quantum transition follows a
rather simple pattern. Nothing, however, can be further from the truth, as we show below.

Consider the interferometric experiment described in Fig.2. The three mirrors used in the experiment are
perfectly reflecting, with mirror M being silvered on both sides. The difference between this arrangement and a
standard interferometer is that one of the beams emerging from the interferometer, does not directly enter the corresponding detector (D$_1$). Rather, it is first reflected by a supplementary mirror onto the mirror M which
then reflects it towards the detector. Mirror M receives therefore two momentum kicks, one from the light inside
the interferometer and one from the beam that reflects on it from the outside. The focus of interest in this
experiment is the momentum received by the mirror M.

The incidence angles of the inside and outside light beams onto M are chosen such that $\cos\beta={1\over2}\cos
\alpha$. The beamsplitters BS$_1$ and BS$_2$ are identical, having reflectivity $r$ and transmissivity $t$, both
given by real numbers with $r^2+t^2=1$, and  $r>t$. Using the standard convention, an incoming
state $|{\rm in}\ra$ impinging on the beamsplitter will emerge as a superposition of a reflected state $|R\ra$ and a transmitted state
$|T\ra$; $|{\rm in}\ra\rightarrow ir|R\ra+t|T\ra$. Hence when a single photon impinges from the left on BS$_1$,
as illustrated in Fig. 2, the effect of the beamsplitter is to produce inside the interferometer the state
\beq|\Psi\ra=ir|A\ra+t|B\ra\eeq where $|A\ra$ and $|B\ra$ denote the photon propagating along the $A$ and $B$
arms of the interferometer respectively.

\begin{figure}
\epsfig{file=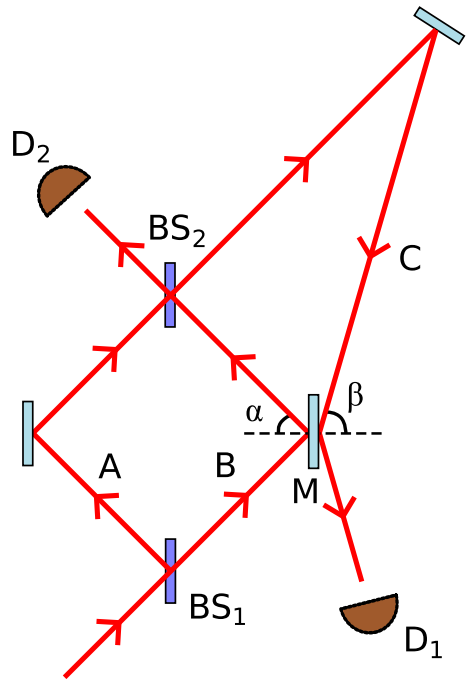, scale=0.4} \caption{Mach-Zehnder interferometer with one output beam reflected back onto the exterior side of mirror M. }
\end{figure}

The second beamsplitter, BS$_2$, is identical to the first. As one can readily check, a photon
in the quantum state $|\Phi_1\ra=t|A\ra-ir|B\ra $ impinging on the second beamsplitter, emerges towards
detector D$_1$ while a photon in the orthogonal state $|\Phi_2\ra=-ir|A\ra+t|B\ra$ emerges towards detector D$_2$.

Thus, when a single photon enters the interferometer by impinging on the left side of the
beamsplitter BS$_1$, the probabilities to be found in the arms $A$ and $B$ are $r^2$ and $t^2$ respectively. The
probability of emerging towards detector D$_1$ is $\big |\la \Phi_1|\Psi\ra\big |^2=4r^2t^2$ while the
probability of emerging towards D$_2$ is $\big |\la \Phi_2|\Psi\ra\big |^2=(r^2-t^2)^2=1-4r^2t^2$.

Suppose now that we send a classical light beam of intensity $I$ towards this interferometer.  Given the above
settings, the intensity of light in arm $A$ is $I_A=r^2I$ while the intensity in arm $B$ is $I_B=t^2I$. The
intensities of the output beams are $I_{D_1}=4r^2t^2I$ and $I_{D_2}=(1-4r^2t^2)I$.

Again, let the beam have a unit area and the experiment last a unit time. To find the momentum given to the
mirror is straightforward. The momentum given by the beam inside the interferometer is
$2I_B\cos\alpha=2t^2I\cos\alpha$. The momentum given by the beam hitting the mirror from the outside is
$2I_{D_1}\cos\beta=8r^2t^2I\cos\beta$. The total momentum is thus \beq\delta p_M=
2t^2I(\cos\alpha-4r^2\cos\beta).\eeq Using the fact that, by construction, $\cos\beta={1\over2}\cos\alpha$ we
obtain
\beq\delta p_M=2t^2I(1-2r^2)\cos\alpha=-2t^2I(r^2-t^2)\cos\alpha.\label{momentumM}\eeq
Since we took
$r>t$, the sign of the momentum received by the mirror is negative, hence the mirror is pushed towards the
inside of the interferometer.

The main result of the experiment is therefore that mirror M receives a net momentum towards the inside of the
interferometer. The calculation above was done using classical optics, but a straightforward quantum calculation
obviously leads to the same result. The issue however is with the story each theory has to tell. According to
the classical optics description of the experiment, the story is very simple. Light impinges on M from two
sides. Although the external beam has a shallower incidence angle than the inside beam, its intensity is much
higher and overall the momentum kick given by it is larger, hence the mirror is pushed inwards. We therefore
conclude that the external beam impinging on M plays the central role - it is due to it that the mirror is pushed
inwards. Hence one would be tempted to assume that quantum mechanically the photons that
constitute this beam are the ones responsible for the inward push. Remarkably, this is not so.

Let us now analyze in detail the momentum given to the mirror by the photons that end up at D$_1$ and at D$_2$. Each photon
incident on M gives it a momentum kick $\delta$. The first thing to notice is that each individual momentum kick is much smaller than the
spread $\Delta$ of momentum of the mirror. This is a general property of any interferometer. It has to be so in order to maintain the coherence of the light in
the interferometer; otherwise the photons will become entangled with the mirror. To show this, we note that a photon when going
through arm A will produce no kick to M while when going through arm B will produce a kick. Accordingly, if $\phi(p)$ is the initial quantum state of the mirror and by $|\Psi\ra$ the quantum state of the photon after the input beamsplitter BS1, but before reaching the mirror, the reflection on the mirror results in

\beq |\Psi\ra\phi(p)=(ir|A\ra+t|B\ra)\phi(p)\rightarrow ir|A\ra\phi(p)+t|B\ra\phi(p-\delta)\label{state}.\eeq
If $\phi(p)$ is orthogonal to $\phi(p-\delta)$ where
$\delta=2\hbar \omega\cos\alpha$ is the kick given by the photon, then the photon ends up entangled with the
mirror and coherence is lost. Another way of looking at this is to note that the mirror has to be
localized within a distance smaller than the wavelength of light, otherwise there will be phase fluctuations
larger than $2\pi$ and interference is lost. (In fact the spread $\Delta$ in the momentum of the mirror has to be many times
bigger - of order $\sqrt {\overline n}$ times - than that of an individual kick to ensure coherence when a beam
with an average of $\overline n$ photons and a spread $\sqrt {\overline n}$ goes through the interferometer. At
the same time $\Delta$, being of order $\sqrt {\overline n}\delta$, is small enough so that the average kick, which is of order
${\overline n}\delta$ is detectable.)

For simplicity we take the state of the mirror to be (up to normalisation)
$\phi(p)=\exp({-{{p^2}\over{2\Delta^2}}})$.  Consider now a single photon propagating through the
interferometer. Given that $\delta\ll\Delta$, we can approximate the state (\ref{state}) of the photon and
mirror just before the photon reaches the output beamsplitter by \barr &&|\Psi\ra\phi(p)\approx
ir|A\ra\phi(p)+t|B\ra\Big(\phi(p)-{{d\phi(p)}
\over{dp}}\delta\Big)\nonumber\\&&=|\Psi\ra\phi(p)-t|B\ra{{d\phi(p)} \over{dp}}\delta.\earr

Suppose now that the photon emerges in the beam directed towards D$_1$. The state of the mirror is then given
(up to normalisation) by projecting the joint state onto the state of the photon corresponding to this beam,
i.e.

\barr&&\la\Phi_1|\Big(|\Psi\ra\phi(p)-t|B\ra{{d\phi(p)} \over{dp}}\delta\Big)\nonumber\\&=&\la
\Phi_1|\Psi\ra\Big(\phi(p)-{{t\la\Phi_1|B\ra}\over{\la \Phi_1|\Psi\ra}}{{d\phi(p)}
\over{dp}}\delta\Big)\nonumber\\&=&\la \Phi_1|\Psi\ra\Big(\phi(p)-{{\la\Phi_1|P_B|\Psi\ra}\over{\la
\Phi_1|\Psi\ra}}{{d\phi(p)} \over{dp}}\delta\Big)\nonumber\\&=&\la
\Phi_1|\Psi\ra\phi(p-P_B^w\delta).\label{mirrorstate}\earr Here $P_B=|B\ra\la B|$ is the projection operator on
state $|B\ra$ and $P_B^w={{\la\Phi_1|P_B|\Psi\ra}\over{\la \Phi_1|\Psi\ra}}$ is the so called weak value of
$P_B$ between the initial state $|\Psi\ra$ and the final state $|\Phi_1\ra$ \cite{weak}. The value of $P_B^w$ is
readily found to be

\barr P_B^w&=&{{\la \Phi_1| P_B|\Psi\ra}\over{\la \Phi_1|\Psi\ra}}={{(t\la A|+ir \la B|)P_B(ir|A\ra+t
|B\ra)}\over{(t\la A|+ir \la B|)(ir|A\ra+t |B\ra)}}\nonumber\\&=&{{rt}\over{tr+rt}}={1\over2}\earr
which leads
to the conclusion that the state of the mirror when the photon emerges from the interferometer is
$\phi(p-{1\over2}\delta)$. In other words, a photon that emerges in the D$_1$ beam changes the mirror by $\delta
p_M={1\over2}\delta=\hbar \omega\cos\alpha$. This momentum change is a result of the mirror receiving a {\it
superposition} between a kick $\delta$ and no kick at all, corresponding to the photon propagating through the
two arms.

The appearance in the above of the weak value of the projector $P_B$ is
not accidental. Indeed, we can view the mirror as a measuring device measuring whether or not the photon is in
arm B or not. The momentum of the mirror acts as a ``pointer'' (no kick - the photon is in arm A; kick - the
photon is in arm B). However, since the photon can only change the position of the pointer (i.e. the momentum
of the mirror) by far less than its spread, we are in the so called ``weak measurement'' regime \cite{weak}.

So far, we discussed the kick given to the mirror by  photons which eventually emerges towards D1, by the collision inside the interferometer. After leaving the interferometer towards D$_1$ the photon gives another kick to the mirror, this
time from the exterior. This kick is simply $2\hbar \omega\cos\beta$. Recalling that
$\cos\beta={1\over2}\cos\alpha$ we reach the crucial point of our paper: the total kick given to the mirror by a
photon that ends up in D$_1$ is \beq \delta p_M=\hbar \omega\cos\alpha-2\hbar \omega\cos\beta=0.\eeq Since the
above experiment is in the linear optics regime, the conclusion reached above for a single photon is valid for
all the photons that end up at D$_1$: none of them gives an overall kick to the mirror.

Then who gives the mirror its momentum? The photons that emerge towards D$_2$. Indeed, the state of the mirror in this case can be computed by a calculation similar to
the one above, by simply replacing $\la \Phi_1|$ by $\la \Phi_2|$.

The state of the mirror can be readily found to be
\barr&&\la\Phi_2|(|\Psi\ra\phi(p)-t|B\ra{{d\phi(p)} \over{dp}}\delta)\nonumber\\&=&\la
\Phi_2|\Psi\ra\phi(p-P_B^w\delta)\earr with $P_B^w$ being in this case the weak value of $P_B$ between
$|\Psi\ra$ and $|\Phi_2\ra$, \barr P_B^w&=&{{\la \Phi_2| P_B|\Psi\ra}\over{\la \Phi_2|\Psi\ra}}={{(ir\la A|+t
\la B|)P_B(ir|A\ra+t |B\ra)}\over{(ir\la A|+t \la B|)(ir|A\ra+t
|B\ra)}}\nonumber\\&=&-{{t^2}\over{r^2-t^2}}.\earr Hence the momentum kick received by the mirror due to a
photon emerging towards D$_2$ is \beq\delta p_M=P_B^w\delta=-{{t^2}\over{r^2-t^2}}2\hbar \omega\cos\alpha.\eeq
Finally, the total momentum given to the mirror by all the photons emerging towards D$_2$ is given by the
momentum due to each photon times the number of photons in the beam. Using the fact that the probability of a
photon to end in this beam is $(r^2-t^2)^2$ we obtain

\beq \delta p_M={\overline n}(r^2-t^2)^2{{-t^2}\over{r^2-t^2}}2\hbar \omega\cos\alpha =-2t^2\hbar
\omega {\overline n}(r^2-t^2)\cos\alpha \eeq identical to the classical result (\ref{momentumM}).

To summarize: The story told by quantum mechanics is dramatically different from the classical one. Classically
it is the light in the D$_1$ beam that is responsible for giving the mirror the kick towards the inside of the
interferometer. Indeed, it is only this beam that can give a momentum in this direction. The light in the
interferometer can either give no momentum (beam A), or a momentum towards the exterior (beam B). It was
therefore tempting to assume that quantum mechanically it is also the case that the photons in the D$_1$ beam
are those responsible for the effect. What we found however is that the photons that end up in the D$_1$ beam
have overall no effect whatsoever - the momentum they give when colliding with the mirror from the outside is
exactly compensated by the momentum they gave the mirror while inside the interferometer, so the total momentum
they give is strictly zero. In fact, according to quantum mechanics, it is the photons that end up in the D$_2$
beam that give the momentum kick. Astonishingly, although they collide with the mirror only from the inside of the
interferometer, they do not push the mirror outwards; rather they somehow succeed to pull it in! This is
realized by a superposition of giving the mirror zero momentum and positive momentum - the superposition results
in the mirror gaining negative momentum.

As mentioned before, our analysis is not limited to interpretational issues. It also serves as a guide for our
intuition. Suppose that, by a quantum fluctuation, we receive more than the average number of photons at
detector D$_1$. The classical intuition will led us to expect that now the mirror will receive an even larger
momentum inwards. Our analysis above tells us differently - since the effect is due to photons going towards
D$_2$ and now there are fewer of them, the inward momentum will be smaller.

{\bf Acknowledgments} This work was supported in part by the Binational Science Foundation Grant 32/08, the Israel
Science Foundation  Grant No. 1125/10, the Templeton Foundation, ERC Advanced grant NLST and EU grant Qessence.

\end{document}